\documentclass[aps,prd,onecolumn, showpacs,floatfix]{revtex4}
\usepackage{amssymb,amsmath,epsfig,graphicx}
\usepackage{caption}
\usepackage{dcolumn}

\newcommand{\be}{\begin{equation}}
\newcommand{\ee}{\end{equation}}
\newcommand{\bea}{\begin{eqnarray}}
\newcommand{\eea}{\end{eqnarray}}
\newcommand{\bi}{\begin{itemize}}
\newcommand{\ei}{\end{itemize}}

\begin{document}

\title{\bf Out of equilibrium quantum field dynamics of an
initial thermal state after a change in the external field}
\author{F. J. Cao}
\email{francao@fis.ucm.es}
\affiliation{Departamento de F\'{\i}sica At\'omica, Molecular y
Nuclear,
Universidad Complutense de Madrid, \\
Avenida Complutense s/n, 28040 Madrid, Spain.}
\affiliation{LERMA, Observatoire de Paris, Laboratoire Associ\'e au CNRS UMR 8112, \\
61, Avenue de l'Observatoire, 75014 Paris, France.}

\author{M. Feito}
\email{feito@fis.ucm.es}
\affiliation{Departamento F\'{\i}sica At\'omica, Molecular y
Nuclear,
Universidad Complutense de Madrid, \\
Avenida Complutense s/n, 28040 Madrid, Spain}
\date{\today}

\begin{abstract}
The effects of the initial temperature in the out of equilibrium
quantum field dynamics in the presence of an homogeneous external
field are investigated. We consider an initial thermal state of
temperature $ T $ for a constant external field $ \vec J $. A
subsequent sign flip of the external field, $ \vec J \to - \vec J
$, gives rise to an out of equilibrium nonperturbative quantum
field dynamics. The dynamics is studied here for the symmetry
broken $\lambda(\vec\Phi^2)^2$ scalar $ N $ component field theory
in the large $ N $ limit. We find a dynamical effective potential
for the expectation value that helps to understand the dynamics.
The dynamics presents two regimes defined by the presence or
absence of a temporal trapping close to the metastable equilibrium
position of the potential. The two regimes are separated by a
critical value of the external field that depends on the initial
temperature. The temporal trapping is shorter for larger initial
temperatures or larger external fields. Parametric resonances and
spinodal instabilities amplify the quantum fluctuations in the
field components transverse to the external field. When there is a
temporal trapping this is the main mechanism that allows the
system to escape from the metastable state for large $ N $.
Subsequently backreaction stops the growth of the quantum
fluctuations and the system enters a quasiperiodic regime.
\end{abstract}

\maketitle

\tableofcontents

\section{Introduction} \label{sec:intro}

Out of equilibrium dense concentrations of particles are present
in several important physical systems, as the ultrarelativistic
heavy ion collisions \cite{ioncol} and the early universe
\cite{tsuinf}. These out of equilibrium dense concentrations of
particles imply the need of out of equilibrium nonperturbative
quantum field theory methods, as the large $N$ limit.

We study here the effects of the initial temperature in the out of
equilibrium dynamics induced by a change in the external field. We
consider the $ O(N) $ $ \lambda \vec\Phi^4 $ theory with
spontaneously broken symmetry. Initially, an homogeneous external
field $ \vec{\cal J} $ breaks the vacuum degeneracy and the system
is in a thermal equilibrium state at a given temperature $ T $.
Subsequently, the external field $ \vec J $ flips the sign
inducing an out of equilibrium dynamics in the system. We study
this out of equilibrium dynamics using the large $ N $ limit
method. We pay a particular attention to the effects of the
initial temperature in the dynamics of the system. This system for
the particular case of zero temperature has been studied in
Ref.~\cite{extfield}. We show here how those results are extended
or modified for nonzero initial temperatures. On the other hand,
the effects of uniform external fields in $ \lambda \vec\Phi^4 $
theory with broken symmetry have been studied in a different
framework through classical evolution with random initial
conditions \cite{clas}. Here we study the out of equilibrium
quantum field dynamics of an initial thermal state after a change
in the external field.

In Section~\ref{sec:model} we present the $\lambda(\vec\Phi^2)^2$
model and its out of equilibrium evolution equations in the large
$ N $ limit. The evolution equation for the expectation value can
be restated in terms of an effective dynamical potential as it is
shown in Section~\ref{sec:effpot}. This effective dynamical
potential and its maxima and minima help to understand the
dynamics. The early time dynamics is presented in
Section~\ref{sec:regimenes}, where we show that there are two
dynamical regimes defined by whether the system is temporally
trapped close to a metastable state or not. The dynamical regime
is determined by the values of the initial temperature and the
external field, because the two regimes are separated by a
critical value of the external field, $J_c(T)$, that depends on
the initial temperature. In this section we also compute the trapping
time finding that it is shorter for larger initial temperatures or
larger external fields. In the next section, Section
\ref{sec:inttime}, we study the intermediate time dynamics and
find a quasiperiodic evolution with a clear separation of slow and
fast variables. Finally, we present the conclusions and an
appendix where the analytic expression for the spinodal time (the
trapping time) is derived as a function of the external field and
the initial temperature.

\section{The model} \label{sec:model}

We consider $N$ scalar fields, $\vec\Phi$, with a $
\lambda(\vec\Phi^2)^2$ selfinteraction in the presence of an
external field $\cal {\vec J}$. The action and the lagrangian
density are given by
\bea
S &=& \int{d^4x {\cal L} }\;, \\
{\cal L} &=& \frac12[\partial_\mu\vec\Phi(x)]^2 - \frac12 m^2
\vec\Phi^2 -\frac{\lambda}{8N}(\vec\Phi^2)^2 -
\frac{m^4N}{2\lambda}+\vec{\cal J} \cdot \vec\Phi\;.\label{modelo}
\eea
We restrict ourselves to the case where the symmetry is
spontaneously broken, i.e., $m^2<0$; and we mainly consider small
coupling constants $ \lambda $, because this slows the dynamics
and allows a better study of its different parts.

We consider here the evolution of a initial thermal state of
temperature $T$ after a flip in the homogeneous external field
$\vec{\cal J}\to -\vec{\cal J}$. A thermal state of temperature
$T$ has translational invariance. This translational invariance of
the initial state is preserved by the evolution equations when
the external field $\vec{\cal J}$ is homogeneous, as in our case.
Thus, the expectation values of the fields, in
particular $\langle\vec\Phi\rangle$ and
$\langle\vec\Phi^2\rangle$, are independent of the spatial
coordinates and only depend on time. We want to study the dynamics
after the change $\vec{\cal J}\to -\vec{\cal J}$; this implies
that the direction of the external field is fixed, and we can
choose the axes in the $N$-dimensional internal space such that
\be
\vec{\cal J} = \left\{
 \begin{array}{lll}
   (\sqrt{N}J,0,\ldots,0) & \mbox{for} & t\leq 0, \\
   (-\sqrt{N}J,0,\ldots,0) & \mbox{for} &t> 0.
 \end{array}
 \right.
\ee
For an initial thermal state we have an expectation value parallel
to the external field. The $O(N)$ invariance of the
lagrangian~\eqref{modelo} for $\vec{\cal J}=0$ together with the
fixed direction of $\vec{\cal J}$ guarantees that the expectation
value remains parallel to the external field during the evolution.
Therefore, the following decomposition can be done
\be
\vec\Phi(x) = \left( \sigma(x), \vec\pi(x) \right) =
\left(\sqrt{N} \phi(t) + \chi(x), \; \vec\pi(x) \right)\;,
\ee
with $ \sqrt{N}\phi(t) = \langle \sigma(x) \rangle $; thus,
$\langle \chi(x) \rangle = 0 $. While in the remaining $ N-1 $
directions transversal to the expectation value, $\langle
\vec\pi(x) \rangle=0$.

\subsection{Evolution equations in the large $N$ limit}
We present here the main concepts and results that leads to the
derivation of the evolution equations in the large $ N $ limit.
More details can be found in Refs.~\cite{extfield,tsu}.
\par
As we have one direction parallel to the expectation value and $N-1$
transversal, the fluctuations in the transverse directions dominate in the
large $ N $ limit, while those in the longitudinal direction only contribute
to the evolution equations as corrections of order $1/N$.
\par
In the Heisenberg picture we have
\be
\vec\pi(\vec x, t) = \int{ \frac{d^3\vec k}{\sqrt{2}(2\pi)^3}
\left[ \vec a_{\vec k} \varphi_{\vec k}(t) e^{i\vec k \cdot \vec
x} + \vec a_{\vec k}^\dagger \varphi_{\vec k}^*(t) e^{-i\vec k
\cdot \vec x} \right] }\;,
\ee
where $\vec a_{\vec k} $, $ \vec a_{\vec k}^\dagger $ are the
annihilation and creation operators respectively, that satisfy the
usual canonical conmutation rules. The functions $\varphi_{\vec
k}(t)$ are the mode functions of the $ \vec\pi $ field.
\par
The evolution equations for $t>0$ are
\begin{eqnarray}\label{evolution1}
 &&\ddot \phi(t) + {\cal M}_d^2(t) \phi(t) = -J \;, \\
\label{evolution2}
 &&\ddot \varphi_{\vec k}(t) + \omega_{\vec k}^2(t) \varphi_{\vec k}(t) = 0 \;,
\end{eqnarray}
where we define the effective frequency $\omega_{\vec k}$ as
\begin{equation}
\omega_{\vec k}^2(t) \equiv k^2 + {\cal M}_d^2(t) \;,
\end{equation}
\begin{equation}
{\cal M}_d^2 \equiv m^2 + \frac{\lambda}{2} \left[ \phi^2(t) +
\frac{\langle \vec\pi^2 \rangle(t)}{N} \right] \;.
\end{equation}
The last term in the effective mass squared, ${\cal M}_d^2$, gives
the quantum and thermal contributions. It can be interpreted as
the mean effect due to the $\vec\pi$ particles, and it is given by
\begin{eqnarray}
&&\frac{\langle \vec\pi ^2 \rangle}{N} = \frac12 \int{
\frac{d^3{\vec k}}{(2\pi)^3} \left[ |\varphi_{\vec
k}(t)|^2\coth\left(\frac{\beta_d\omega_{\vec k}(0)}{2}\right) -
{\cal S}_d \right] } \;,\label{pi2} \\
&&{\cal S}_d \equiv \frac1k - \frac{\theta(k-\kappa)}{2k^3} {\cal
M}_d^2(t) \;,
\end{eqnarray}
with $\beta_d^{-1}=k_B T$, and $\kappa$ an arbitrary
renormalization scale that we choose, for simplicity, equal to the
renormalized mass, $\kappa=|m_R|$. (For details on the
renormalization procedure, that leads to the subtraction ${\cal
S}_d$ see Ref.~\cite{boy}.) The previous equations are already
written for the renormalized magnitudes. It is important to stress
that the renormalization is independent of the
temperature~\cite{kapusta}. The hyperbolic cotangent factor in the
Eq.~\eqref{pi2} is due to the initial thermal state (see for
example Ref.~\cite{feynStatMech}). In addition, the fact that the
initial state is a thermal state implies the following initial
conditions for the expectation values and the modes:
\begin{eqnarray}
&& \phi(0) = \phi_0 \;, \quad \dot\phi(0)=0 \;,\label{ci1} \\
&& \varphi_{\vec k}(0) = \frac{1}{\sqrt{\omega_k(0)}} \;, \quad
\dot\varphi_{\vec k}(0) = - i \sqrt{\omega_k(0)}\label{ci2},
\end{eqnarray}
where $\phi_0$ is the stationary solution for $t\leq 0$ with
minimal total energy. The evolution equation for $t\leq 0$ is the
Eq.~\eqref{evolution1} changing $J$ for $-J$, therefore the
stationarity condition for $t\leq 0$ is ${\cal M}_d^2 \phi_0 = J$.
Note that the initial conditions for the modes are spherically
symmetric in the momentum space and the evolution keeps this
symmetry (because the evolution equations are also spherically
symmetric).
\par
We introduce the following adimensional variables:
\begin{gather}
\tau\equiv|m|t\;,\quad \eta(\tau)\equiv \sqrt{\frac{\lambda}{2}}\frac{\phi(t)}{|m|}\;,\quad
{\vec \jmath}\equiv\sqrt{\frac{\lambda}{2N}}\frac{{\vec{\cal J}}}{|m|^3}\;,\\
q\equiv\frac{k}{|m|}\;,\quad g\equiv\frac{\lambda}{8\pi^2}\;,\quad
{\cal M}(\tau)\equiv\frac{{\cal M}_d(t)}{|m|}\quad \beta\equiv \frac{\beta_d}{|m|}\;,\\
\varphi_q(\tau)\equiv\sqrt{|m|}\varphi_k(t)\;,\quad
g\Sigma(\tau)\equiv\frac{\lambda}{2|m|^2} \frac{\langle \vec\pi^2
\rangle(t)}{N}\;,\quad {\cal S}\equiv\frac{{\cal S}_d}{|m|}\;,
\end{gather}
in terms of these adimensional variables the evolution equations for $\tau>0$
are
\begin{equation}\label{evolution1_adim}
\ddot \eta(\tau) + {\cal M}^2(\tau) \eta(\tau) = -j \;, \quad
\ddot \varphi_q(\tau) + \omega_q^2(\tau) \varphi_q(\tau) = 0 \;,
\end{equation}
where
\begin{eqnarray}
&&{\cal M}^2(\tau)=-1+\eta^2(\tau)+g\Sigma(\tau), \quad
\omega_q^2=q^2+{\cal M}^2(\tau)\;,\label{frec_t} \\
&&g\Sigma(\tau)=g
\int{q^2 dq \left[|\varphi_q(\tau)|^2\coth\left(\frac{\beta\omega_q(0)}{2}\right)
-{\cal S}\right]}\;, \\
&&{\cal S}(\tau) \equiv \frac1q - \frac{\theta(q-1)}{2q^3} {\cal
M}^2(\tau) \label{S}\;.
\end{eqnarray}
The initial conditions~\eqref{ci1} and~\eqref{ci2} are expressed as
\begin{eqnarray}
&& \eta(0) = \eta_0 \;, \quad \dot\eta(0)=0 \;,\label{ci_adim1} \\
&& \varphi_q(0) = \frac{1}{\sqrt{\omega_q(0)}} \;, \quad
\dot\varphi_q(0) = - i \sqrt{\omega_q(0)}\label{ci_adim2}
\end{eqnarray}

\section{Effective dynamical potential for the expectation value}
\label{sec:effpot}

We can define an (adimensionalized) energy for positive times as
the (adimensionalized) expectation value of the $T^{00}$ component
of the energy-momentum tensor.
\begin{equation}\label{energia}\begin{split}
\epsilon(\tau)&\equiv\frac{\lambda}{2N|m|^4}\langle T^{00}\rangle=
\frac{\eta^4}{4}-\frac{\eta^2}{2}+\frac{1}{2}\eta^2
g\Sigma-\frac{g\Sigma}{2}+\frac{(g\Sigma)^2}{4}\\
&\quad+\frac{1}{4}+j\eta+\frac{\dot\eta^2}{2}+
\frac{g}{2}\int q^2dq\left[|\dot\varphi_q|^2
\coth{\left(\frac{\beta\omega_q(0)}{2}\right)}-{\cal S}_1\right]\\
&\quad+\frac{g}{2}\int q^2dqq^2\left[|\varphi_q|^2
\coth{\left(\frac{\beta\omega_q(0)}{2}\right)}-{\cal
S}_2\right]\;;
\end{split}\end{equation}
where ${\cal S}_1$ and ${\cal S}_2$ are two subtractions due to
the renormalization
\begin{equation}\begin{split}
{\cal S}_1 &\equiv q + \frac{{\cal M}^2}{2q} - \frac{\theta(q-1)}{8q^3}
 \left[\left({\cal M}^2\right)^2 + \frac{d^2{\cal M}^2}{dt^2} \right]\;,\\
{\cal S}_2 &\equiv \frac1q - \frac{{\cal M}^2}{2q^3} +
\frac{\theta(q-1)}{8q^5}\left[ 3 \left({\cal M}^2\right)^2 + \frac{d^2{\cal
M}^2}{dt^2} \right]\;.
\end{split}\end{equation}
For a time independent external field $j$, as it is the case for
$\tau>0$, the energy is conserved. From the
expression~\eqref{energia} we have in the $(\eta,g\Sigma)$ plane
the restriction $\epsilon\geq V_{de;\tau > 0}(\eta, \Sigma)$,
where
\be
V_{de;\tau > 0}(\eta, \Sigma) \equiv \frac{\eta^4}{4} -
\frac{\eta^2}{2} + \frac12\eta^2 g\Sigma - \frac{g\Sigma}{2} +
\frac{(g\Sigma)^2}{4} + \frac14+j\eta\;.
\ee
$V_{de;\tau>0}(\eta, \Sigma)$ can be interpreted as a dynamical
effective potential because the evolution
equation~\eqref{evolution1_adim} for $\eta$ can be written as
\begin{equation}
\ddot\eta(\tau)=-\frac{\partial}{\partial \eta}V_{de;\tau>0}(\eta, \Sigma)\;.
\end{equation}
It must be stressed that $V_{de;\tau>0}$ is an effective potential
\emph{only} for $\eta$ (and \emph{not} for the modes).

\subsection{Equilibrium states for the effective dynamical potential}\label{sec:estados}
\begin{figure}
\begin{center}
\includegraphics [scale=0.7] {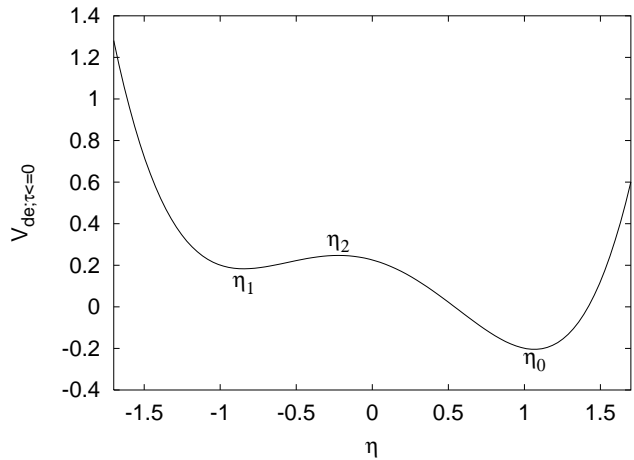}
\end{center}
\caption{Dynamical effective potential $V_{de;\tau\leq 0}$ as a
function of $\eta$ for $g\Sigma(0)=0.05$. The value of the
external field is $j=0.2$. The positions of the global minimum
$\eta_0$, the local minimum $\eta_1$ (metastable state), and the
relative maximum $\eta_2$ are shown. \label{potenciales3}}
\end{figure}

The dynamical effective potential for $\tau\leq 0$ can be defined
as $V_{de;\tau\leq 0}(\eta, \Sigma) = V_{de;\tau>0}(\eta,\Sigma)
-2j\eta$. Therefore, the stationary states for the initial
dynamics for times $\tau\leq 0$ (before the external field has
been flipped) are the solutions of $V_{de;\tau\leq 0}'(\eta) = 0$
(the apostrophe means $\eta$ derivative). Thus, they verify
\be \label{statstates}
\eta^3+(-1+g\Sigma)\eta -j =0\;.
\ee
For small external fields,
\be
j<j_d \equiv 2\sqrt{\frac{(1-g\Sigma)}{27}},
\ee
we have three roots. There is a global minimum, that corresponds
to a stable equilibrium state, at the value of $ \eta $
\be\begin{split} \label{eta0}
\eta_0 &\equiv 2\sqrt{\frac{1-g\Sigma}{3}}\cos\left\{
\frac13 \arccos\left[ \frac{j}{2}\sqrt{\frac{27}{(1-g\Sigma)^3}}
\right] \right\}\\
&=(1-g\Sigma)^{1/2}+\frac{1}{2}(1-g\Sigma)^{-1}j-\frac{3}{8}
(1-g\Sigma)^{-5/2}j^2\\
&\quad+\frac{1}{2}(1-g\Sigma)^{-4}j^3-\frac{105}{128}(1-g\Sigma)^{-11/2}j^4
+{\cal O}(j^5)
\;,
\end{split}\ee
a local minimum (metastable equilibrium state) at
\be\begin{split}
\eta_1 &\equiv 2\sqrt{\frac{1-g\Sigma}{3}}\cos\left\{
\frac13 \arccos\left[ \frac{j}{2}\sqrt{\frac{27}{(1-g\Sigma)^3}}
\right] +\frac{2\pi}{3}\right\}\\
&=-(1-g\Sigma)^{1/2}+\frac{1}{2}(1-g\Sigma)^{-1}j+\frac{3}{8}
(1-g\Sigma)^{-5/2}j^2\\
&\quad+\frac{1}{2}(1-g\Sigma)^{-4}j^3+\frac{105}{128}(1-g\Sigma)^{-11/2}j^4
+{\cal O}(j^5)
\;,
\end{split}\ee
and a local maximum (unstable equilibrium) at
\be\begin{split}
\eta_2 &\equiv 2\sqrt{\frac{1-g\Sigma}{3}}\cos\left\{
\frac13 \arccos\left[ \frac{j}{2}\sqrt{\frac{27}{(1-g\Sigma)^3}}
\right] +\frac{4\pi}{3}\right\}\\
&= -(1-g\Sigma)^{-1}j-(1-g\Sigma)^{-4}j^3+{\cal O}(j^5)
\;.
\end{split}\ee
On the other hand, for large external fields ($j>j_d$) there is a
single extreme, that corresponds to a global minimum (stable
equilibrium), at the value
\be \label{eta0onemin}\begin{split}
\eta_0 &\equiv \left( \frac{j}{2} \right)^{1/3}  \Bigg\{
\left[ 1 + \sqrt{ 1 - \frac{ 4(1-g\Sigma)^3 }{27j^2} }
\right]^{1/3} + \left[ 1 - \sqrt{ 1 - \frac{ 4(1-g\Sigma)^3
}{27j^2} } \right]^{1/3} \Bigg\}\\
&=j^{1/3}+\frac{1}{3}(1-g\Sigma)j^{-1/3}-\frac{1}{81}(1-g\Sigma)^{3}j^{-5/3}
+{\cal O}(j^{-7/3})
\;.
\end{split}\ee
We consider here the more interesting case, $j<j_d$, where a
potential barrier is present. The initial conditions are those of
the stable equilibrium state of the dynamics for times $\tau\leq
0$. In Fig.~\ref{potenciales3} we show the location of the minima
and the maximum of $V_{de;\tau\leq 0}(\eta)$ for $g\Sigma(0)=0.05$
and an external field $j=0.2<j_d$.

For zero initial temperature $ g\Sigma(0) \sim g \ll 1 $. Thus,
the zero temperature results of Ref.~\cite{extfield} are recovered
neglecting $ g\Sigma $ in Eqs.~\eqref{statstates}-\eqref{eta0onemin}.

\section{Early time dynamics and dynamical regimes}\label{sec:regimenes}

The fact that the initial state is not the ground state but a
thermal state increases the initial value of $ g \Sigma $. It can
be shown that $ \Sigma(0) $ is approximately given by
\be \label{SigmaT}
\Sigma(0) = \Sigma^{T=0}(0) + \frac{\pi^2}{3} \beta^{-2}\;.
\ee
$\Sigma^{T=0}(0)$ is the zero temperature value, that for $g\ll 1$
only depends on $j$ (see Table~\ref{D1}). The
second term on the righthand side of Eq.~\eqref{SigmaT} is the
thermal contribution computed in the hard thermal loop
approximation \cite{lebellac}, i.e., assuming the main contribution
comes from the modes with $ q \sim \beta^{-1} $ and we are in the case $
\beta^{-1} \gg {\cal M}^2(0) $. [On the other hand, for
$ \beta^{-1} \ll {\cal M}^2(0) $ thermal effects are negligeable,
because $ \beta^{-1} \ll \omega_q $ and $ \coth(\beta\omega_q/2) =
1 + {\cal O}(e^{-\beta\omega_q})$.]

\bigskip

\begin{table}[h]
\begin{tabular}{c|ccccc}
\hline \hline
$j$ & $0.01$ &$0.05$ & $0.10$ & $0.20$&$0.25$ \\ \hline
$-10^{2} \Sigma^{T=0}(0)$ &$1.2$& $4.2$ & $6.5$ & $9.6$&$11$ \\
\hline \hline
\end{tabular}
\caption{ $\Sigma^{T=0}(0) $ values obtained for various values of
the external field $j$ ($g\ll 1$).\label{D1}}
\end{table}

\begin{figure}
\begin{center}
\includegraphics [scale=0.7] {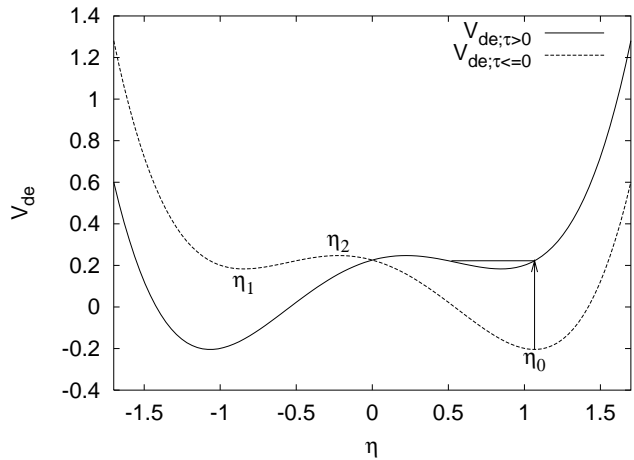}
\end{center}
\caption{Dynamical effective potential $V_{de;\tau\leq 0}$ and
$V_{de;\tau> 0}$ as a function of $\eta$ for $g\Sigma(0)=0.05$.
The value of the external field is $j=0.2$. The positions of the
global minimum $\eta_0$, the local minimum $\eta_1$, and the local
maximum $\eta_2$ are shown. There is a potential barrier ($ j <
j_d $), and the systems gets temporally trapped ($ j < j_c$).
\label{potenciales1}}
\end{figure}

\begin{figure}
\begin{center}
\includegraphics [scale=0.7] {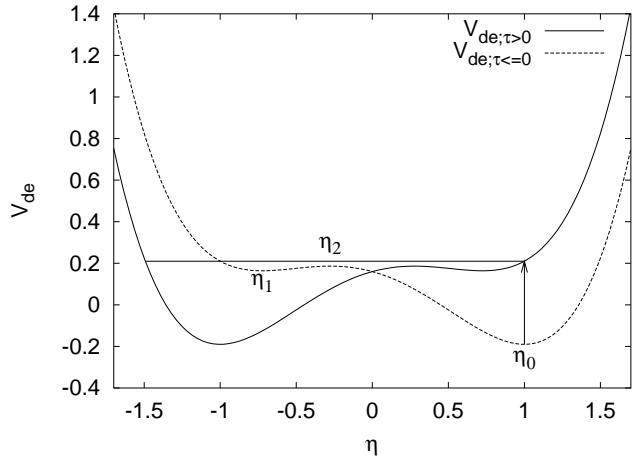}
\end{center}
\caption{Dynamical effective potentials $V_{de;\tau\leq 0}$ and
$V_{de;\tau> 0}$ as a function of $\eta$ for $g\Sigma(0)=0.2$. The
value of the external field is $j=0.2$. The positions of the
global minimum $\eta_0$, the local minimum $\eta_1$ and the local
maximum $\eta_2$ are shown. The potential barrier is not high
enough ($ j < j_c $), and the system rapidly evolves towards the
absolute minimum. \label{potenciales2}}
\end{figure}

After the initial flip of the external field sign at $ \tau=0 $
the positions of the absolute minimum and the relative minimum of
the potential are interchanged, and the state of the system
becomes a metastable state (Figs.~\ref{potenciales1}
and~\ref{potenciales2}). The height of the potential barrier is
$V_{de;\tau > 0}(\eta_2')$, where $\eta_2'=-\eta_2$ is the maximum
of $V_{de;\tau>0}(\eta)$. The existence of this potential barrier
gives rise to two different dynamical regimes. In the first one
the system can directly overcome the barrier
$V_{de;\tau>0}(\eta_0)>V_{de;\tau>0}(\eta_2')$, and rapidly
reaches the neighborhoods of the global minimum. While in the
second regime the system can not overcome the barrier directly
$V_{de;\tau>0}(\eta_0)<V_{de;\tau>0}(\eta_2')$, and it gets
temporally trapped close to the metastable state. Solving the
equation
\be\label{jcdefinida}
V_{de;\tau>0}(\eta_0)=V_{de;\tau>0}(\eta_2')
\ee
the external field critical value $j_c$ that separates the two
regimes, untrapped $|j|>j_c$, or trapped $|j|<j_c$, is obtained
\be
j_c = \frac{j_c^{T=0}}{(1-g\Sigma(0))^{3/2}} \quad \mbox{ with }
\quad j_c^{T=0} = \sqrt{ 2 \; {\left( 13^2 + 15 \sqrt{5} \right)
\over 19^3}} = 0.243019\ldots \;;
\ee
where we have used that for very early times we can make the
approximation $ g\Sigma(\tau)\approx g\Sigma(0) $. An explicit
analytic expression for the critical external field as a function
of the initial temperature is obtained using Eq.~\eqref{SigmaT},
\be \label{jcbeta}
j_c(\beta) = \frac{j_c^{T=0}}{\left[1-
\left(\frac{\beta_c^{J=0}}{\beta}\right)^{2}\right]^{3/2}} \quad
\mbox{ with } \quad \beta_c^{J=0} = \pi\sqrt{\frac{g}{3}} \;.
\ee
This analytical formula is compared with the numerical results for
$g=10^{-2}$ and $g=10^{-3}$ in Fig.~\ref{figjcbeta} finding a good fit.
The two dynamical regimes (untrapped $|j|>j_c$ and trapped
$|j|<j_c$) correspond to the two regions in the $(j,\beta^{-1})$
plane separated by the $j_c(\beta)$ curve. This curve can be
better understood recalling the effects of the temperature and of
the external field on the potential. As the temperature increases
$g\Sigma$ grows, and the barrier in the $\eta$ direction between
the two minima of the potential diminishes until it disappears.
The barrier disappears even for $j=0$ when $ \beta^{-1} >
(\beta_c^{J=0})^{-1} $. On the other hand, the effect of the
external field $j$ is to tilt the potential; for $ j > j_c^{T=0} $
the potential is so tilted that there is no temporal trapping even
for $ \beta^{-1} = 0 $. Therefore, Eq.~\eqref{jcbeta} states how
the trapping can disappear due to a combination of both effects.
\par
We analyze now the two dynamical regimes:

\begin{figure}
\begin{center}
\includegraphics [scale=0.7] {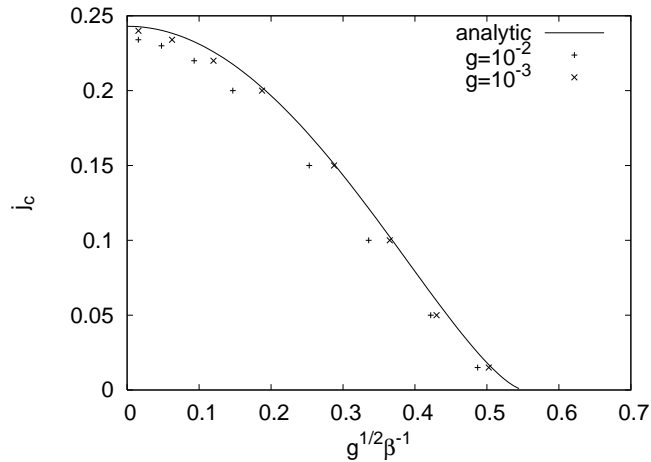}
\end{center}
\caption{Critical external field $j_c$ as a function of
$\beta^{-1}$ for $g=10^{-2}$ and $ g=10^{-3}$ obtained from
numerical simulations (dots) and from the analytical formula in
Eq.~\eqref{jcbeta} (full line). There are two regions in the $(j,\beta^{-1})$
plane corresponding to the two dynamical regimes: for $j<j_c$ the
system is temporally trapped in the metastable state; for $j>j_c$
the system rapidly reaches the neighborhood of the global minimum
of the potential. \label{figjcbeta}}
\end{figure}

\subsection{$j<j_c(\beta)$}\label{atrapado}
\begin{figure}
\begin{center}
\includegraphics [scale=0.7] {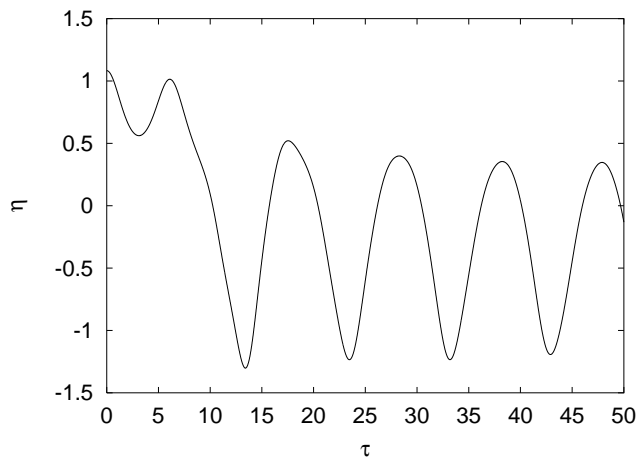}
\end{center}
\caption{$\eta$ as a function of time for $j<j_c(\beta)$. It
corresponds to the values $g=10^{-3}$, $j=0.2$, $\beta^{-1}=2$.
\label{regimenes_eta1}}
\end{figure}

In this regime (Fig.~\ref{potenciales1}) the system does not have
enough energy to jump over the potential barrier and it remains
some time oscillating around the metastable minimum
(Fig.~\ref{regimenes_eta1}). The system is only
\emph{temporally trapped} because the spinodal instability of the
modes (due to the fact that $ {\cal M}^2(\tau) < 0 $) makes the
amplitude of the modes grow. This growth allows the field to
surround the maximum of the potential in the $N-1$ directions
transversal to the external field, and finally the system reaches
the neighborhood of the stable minimum. (Another effect that leads
the system to the global minimum is tunnelling. However, our
computation neglects this effect because it only takes place
in one of the $ N $ internal directions and thereby it is subleading in the
large $ N $ limit.)
\par
The classical evolution equations (recovered for $\Sigma=0$)
predict that the expectation value $\eta$ oscillates forever
between the value $\eta_0$ and the turning point value $\eta_r$
given by
\begin{equation} \label{eqetar}
V_{de;\tau>0}(\eta_r,\Sigma=0)=V_{de;\tau>0}(\eta_0,\Sigma=0).
\end{equation}
This is also the case when $ g\Sigma(0) \ll 1 $ (otherwise the trapping
is very short). Solving Eq. \eqref{eqetar} order by order gives
\begin{equation} \label{etar}
\eta_r=1-\frac{3}{2}j-\frac{11}{8}j^2-\frac{7}{2}j^3-\frac{1049}{128}j^4+{\cal
O}(j^5).
\end{equation}

\begin{figure}
\begin{center}
\includegraphics [scale=0.7] {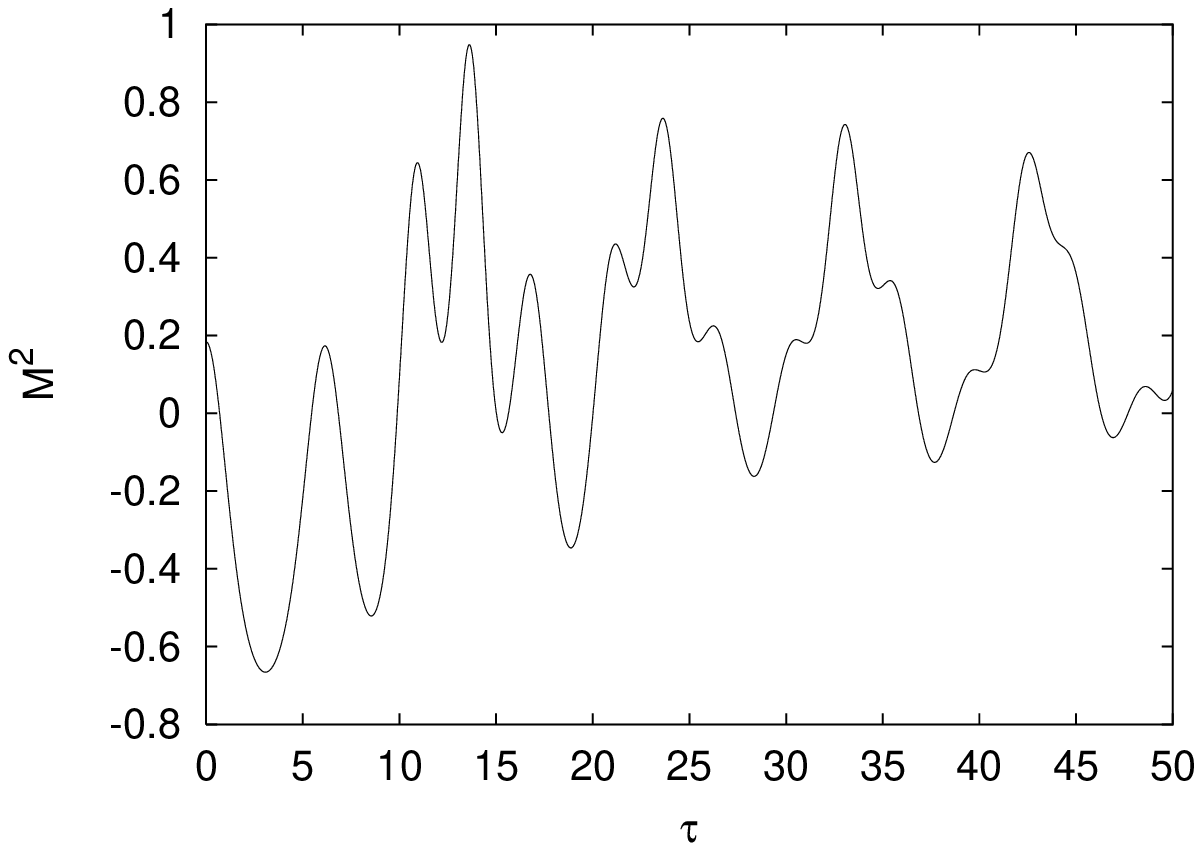}
\end{center}
\caption{${\cal M}^2$ as a function of time for
$j<j_c(\beta)$. It corresponds to the values $g=10^{-3}$,
$j=0.2$, $\beta^{-1}=2$. \label{regimenes_massoft1}}
\end{figure}
\par
\begin{figure}
\begin{center}
\includegraphics [scale=0.7] {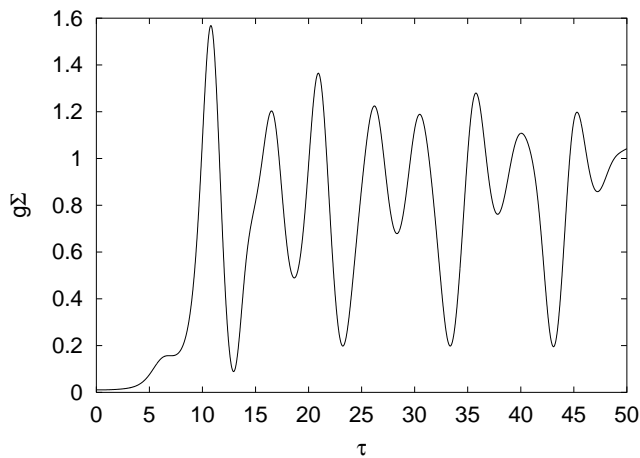}
\end{center}
\caption{$g\Sigma$ as a function of time for $j<j_c(\beta)$.
It corresponds to the values $g=10^{-3}$, $j=0.2$, $\beta^{-1}=2$.
\label{regimenes_gsigma1}}
\end{figure}

In Fig.~\ref{regimenes_massoft1} we plot the behavior of the
squared effective mass. For early times, ${\cal M}^2$ oscillates
with a negative average value. This makes the low momentum modes
spinodally unstable and they grow exponentially. This implies a
quasiexponential growth of $g\Sigma$ for early times, and finally
$g\Sigma$ becomes of order one (see Fig.~\ref{regimenes_gsigma1}).
This makes that $\eta$ is no longer trapped close to the
metastable minimum (see Fig.~\ref{regimenes_eta1}).

Parametric resonances are also present on the evolution, due to
the oscillations of ${\cal M}^2$, implying the transfer of energy
to certain momenta that grow in amplitude. This two mechanisms,
spinodal instability and parametric resonance, transfer energy
from the expectation value $\eta$ to the modes $\varphi_q$
increasing $g\Sigma$. However in this case, $ |j|<j_c $, the main
contribution is the spinodal instability.
\par
A detailed analysis of the spinodal instability allows to obtain
the trapping time, or spinodal time $\tau_s$. (Here we summarize
the results, the details can be found in
Appendix~\ref{tiempoespinodal}.) At early times ($ \tau < \tau_s
$), the effective squared mass, ${\cal M}^2$, oscillate with a
negative average. Thus, we can define an effective average squared
mass $ - \mu^2 $, and in the modes evolution
equation~\eqref{evolution1_adim} replace ${\cal M}^{2}(\tau)$ by
$-\mu^2$ to obtain an estimate for $|\varphi_q|$. This is a good
approximation after several complete oscillations of ${\cal
M}^{2}(\tau)$ (i.e., $ \tau > 1/\mu $), and when the time
dependence of ${\cal M}^{2}(\tau)$ is slow (adiabatic
approximation). The effective average squared mass is given by
[Eq.~\eqref{-mu2enj}]
\begin{equation} \label{-mu2enjmain}
-\mu^2=-j-\frac{1}{2}j^2-\frac{9}{8}j^3-\frac{5}{2}j^4+{\cal
O}(j^5).
\end{equation}
The quasiexponential growth of $ g\Sigma(\tau) $ for $ \tau <
\tau_s $ is
\begin{equation}  \label{gSigmaTmain}
g\Sigma_s(\tau)\approx \left\{
 \begin{array}{lcc}
   \frac{2}{\beta\mu}g\Sigma_s^{T=0}(\tau) & \mbox{for} &
      \beta^{-1} \gg \mu \;,\\
    g\Sigma_s^{T=0}(\tau) & \mbox{for} &
      \beta^{-1} \ll \mu \;,
 \end{array}
 \right.
\end{equation}
with
\be \label{gSigmaT0main}
g\Sigma_s^{T=0}(\tau)\approx\frac{g\sqrt{\pi\mu}e^{2\tau\mu}}{8\tau^{3/2}}\;
\ee
the value for zero temperature.

After a certain time, the spinodal time $\tau_s$, the quantum and
thermal effects start to be important in the dynamics,
$g\Sigma_s(\tau_s)$ compensates $-\mu^2$ and the exponential
growth of the mode functions stops, then the mode functions start to
have an oscillatory behavior. Thus, the spinodal time $\tau_s$ is defined
as
\begin{equation} \label{tsmain}
g\Sigma_s(\tau_s)=\mu^2\;.
\end{equation}
A good approximate expression for the spinodal time is
\be \label{tsfinal}
\tau_s = \left\{
  \begin{array}{lcc}
    \frac{1}{2\mu} \left[\log\left(\frac{8}{g\sqrt{\pi}}\right)
    - \log\left(\frac{2}{\beta\mu}\right) + \frac{3}{2}
    \log(\mu\tau_s) \right]
    & \mbox{ for } & \beta^{-1} \gg \mu \;,\vspace{1 mm}\\
    \frac{1}{2\mu} \left[\log\left(\frac{8}{g\sqrt{\pi}}\right)
    + \frac{3}{2} \log(\mu\tau_s) \right]
    & \mbox{for} & \beta^{-1} \ll \mu \;.\\
  \end{array}
\right.
\ee
We see that the spinodal time for $\beta^{-1} \ll \mu$ is the same
as that for zero temperature $ \tau_s^{T=0} $ (see
Ref.~\cite{extfield}). On the other hand, for $\beta^{-1} \gg \mu$
we see that increasing the initial temperature the spinodal time
decreases. Therefore, a higher initial temperature implies a
shorter trapping period, that is consistent with the fact that for
greater temperatures the initial transversal fluctuations are
larger and the system is closer to the end of the trapping period.
The contribution of the transversal fields grows rapidly following
Eqs.~\eqref{gSigmaTmain}-\eqref{gSigmaT0main}, turning around the
maximum of the potential and reaching the neighborhood of the
absolute minimum at earlier times.
Increasing the external field $j$ also shortens the trapping period,
as can be shown in Eq.~\eqref{tsfinal} ($ \mu \sim \sqrt{j} $).

\begin{table*}
\begin{center}
\begin{tabular}{c|c c c c c c}
\hline\hline
$\frac{\beta}{g}$ & 2$\cdot 10^{2}$ &
2$\cdot 10^{3}$ &
2$\cdot 10^{4}$ & 2$\cdot 10^{5}$ & 2$\cdot 10^{6}$ &
2$\cdot 10^{7}$\\
\hline
$\tau_s (\mathrm{numerical})$ &4.8&9.6&11.5&15.1&16.7&20.1\\
$\tau_s (\mathrm{analytical})$ &7.1&10.1&12.8&15.6&18.2&20.8\\
\hline
$\mathrm{error}\;(\%)$&        33.1 &4.1 &10.8 &3.0 &8.4&3.6\\
\hline\hline
\end{tabular}
\end{center}
\caption{Spinodal time values for different values of
$\frac{\beta}{g}$ with fixed $\beta=0.2$ and $j=0.2$. The values
has been obtained from the numerical simulation of the complete
evolution equations, and from the analytical approximation
Eq.~\eqref{tsfinal}. \label{tabla}}
\end{table*}

In Table~\ref{tabla} we compare the numerical results obtained
solving the evolution equations~\eqref{evolution1}--\eqref{ci2}
and applying the $\tau_s$ definition in Eq.~\eqref{tsmain}, with the
values predicted by the analytical expression in
Eq.~\eqref{tsfinal}. We see a good agreement (with discrepancies of
$10\%$ or smaller) except for low $\frac{\beta}{g}$ values (that
correspond to high temperatures). This later discrepancy is not
troublesome because the expression~\eqref{tsfinal} has been
obtained for times $\tau\gtrsim\frac{1}{\mu}$ and this condition
does not hold for high temperatures, as the the spinodal time
decreases for decreasing $\frac{\beta}{g}$. In fact, when $j=0.2$
we have $\mu^2\approx0.23$ [Eq.~\eqref{-mu2enjmain}] and therefore
$\frac{1}{\mu}\approx 2.1$; as for $\frac{\beta}{g}=2\cdot 10^{2}$
we have a spinodal time $ \tau_s^{\mathrm{sim}} \approx 4.8 \sim
2.1 $ (Table~\ref{tabla}), Eq.~\eqref{tsfinal} can no longer be
applied. However, for greater spinodal times (that corresponds to
$\frac{\beta}{g}\geq 2\cdot 10^{3}$) the behavior predicted by
the Eq.~\eqref{tsfinal} is reproduced with a good agreement,
as we show in Table~\ref{tabla}.

\subsection{$j>j_c(\beta)$}

The main characteristic of this regime (Fig.~\ref{potenciales2})
is that the system has enough energy to overcome the potential
barrier directly. Therefore \emph{no} trapping is present in this
case.

\begin{figure}
\begin{center}
\includegraphics [scale=0.7] {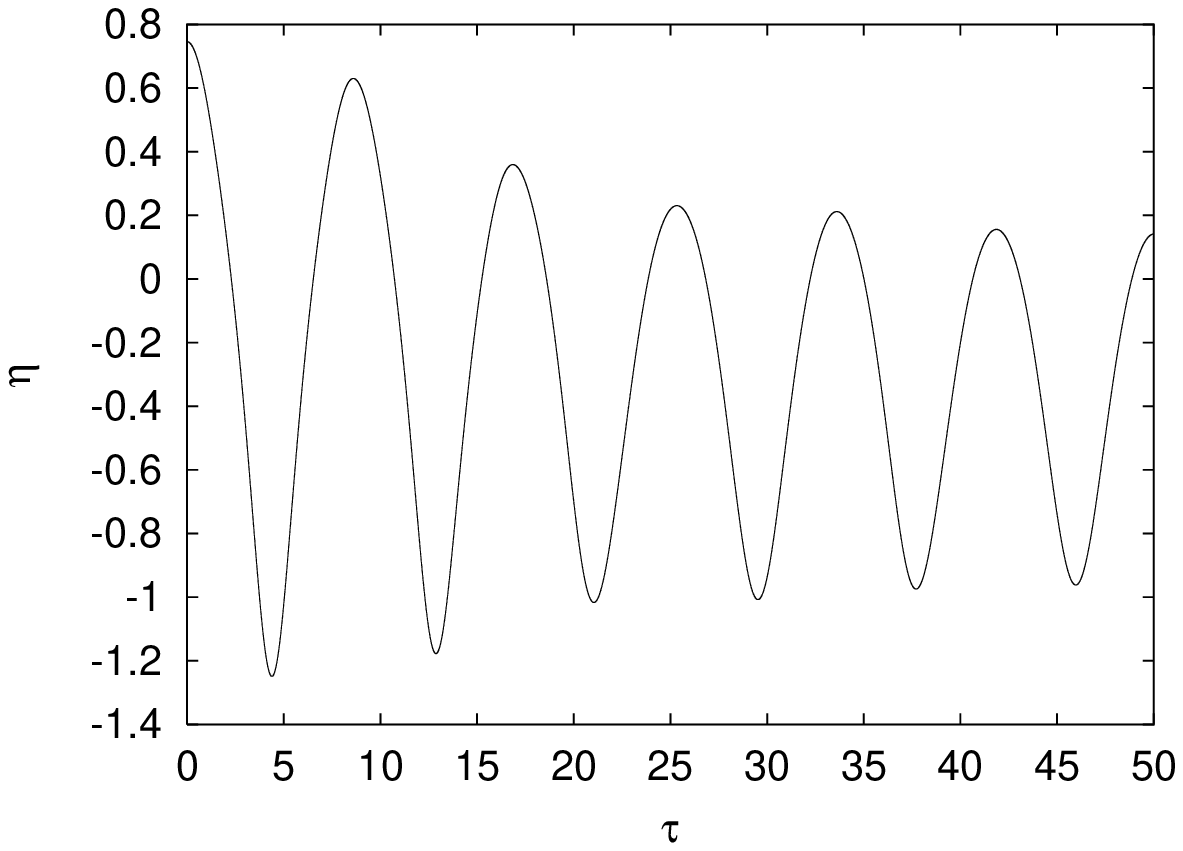}
\end{center}
\caption{$\eta$ as a function of time for $j>j_c(\beta)$.
It corresponds to the values $g=10^{-3}$, $j=0.2$, $\beta^{-1}=15$.
\label{regimenes_eta2}}
\end{figure}

\begin{figure}
\begin{center}
\includegraphics [scale=0.7] {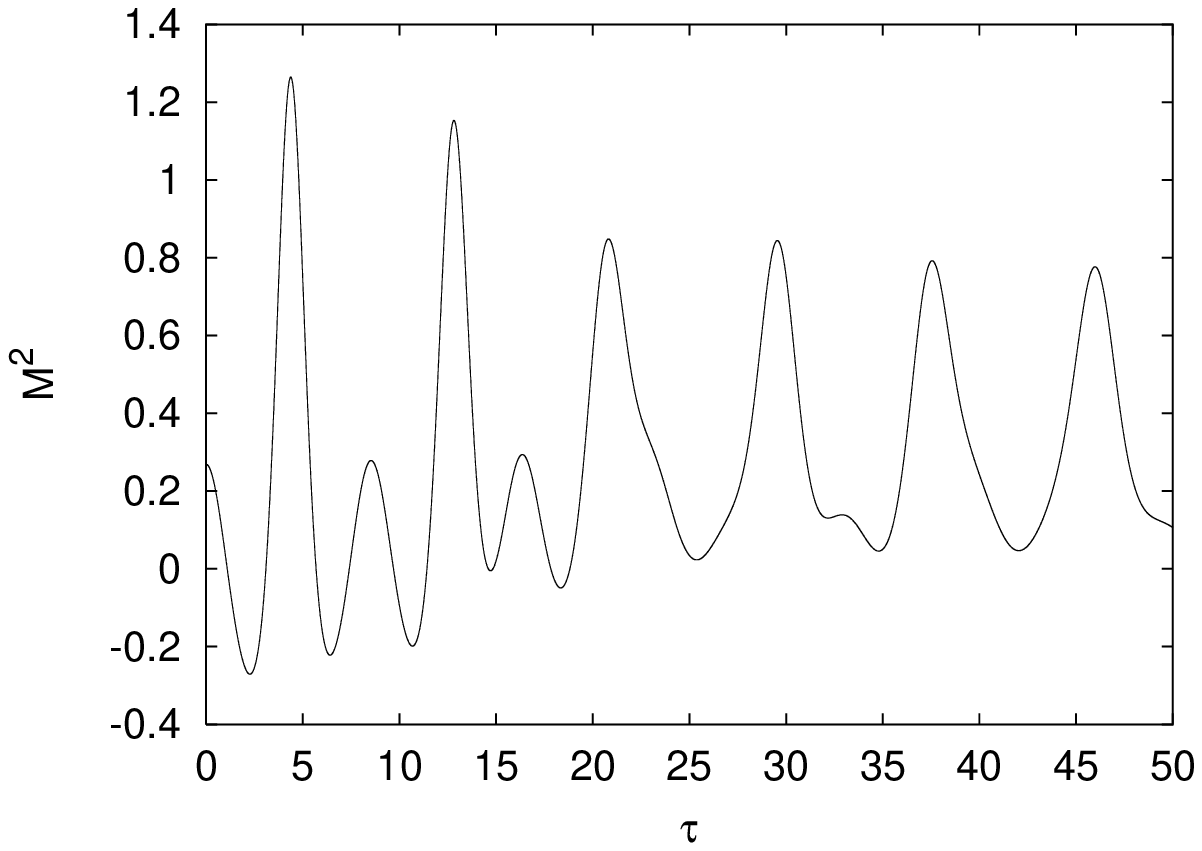}
\end{center}
\caption{$\eta$ as a function of time for $j>j_c(\beta)$.
It coorresponds to the values $g=10^{-3}$, $j=0.2$, $\beta^{-1}=15$.
\label{regimenes_massoft2}}
\end{figure}

\begin{figure}
\begin{center}
\includegraphics [scale=0.7] {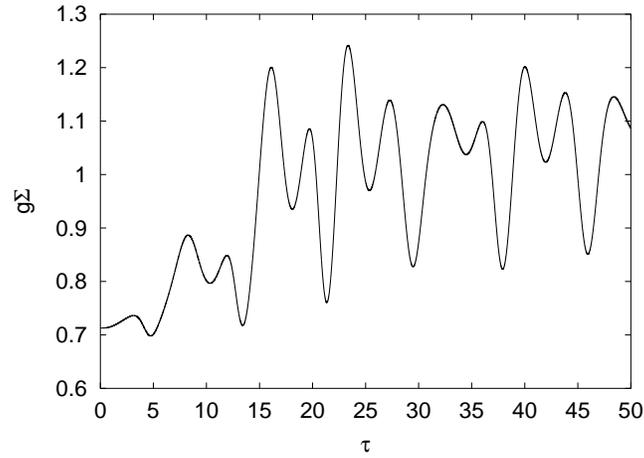}
\end{center}
\caption{$g\Sigma$ as a function of time for $j>j_c(\beta)$.
It corresponds to the values $g=10^{-3}$, $j=0.2$, $\beta^{-1}=15$.
\label{regimenes_gsigma2}}
\end{figure}

Also in this case there is energy transfer between the expectation
value and the modes of the field. The low momenta modes are
spinodally unstable and grow exponentially only during the time
intervals when ${\cal M}^2(\tau)<0$
(Fig.~\ref{regimenes_massoft2}). Parametric resonances are also
present due to the oscillations of the effective squared mass. The
Figs.~\ref{regimenes_eta2},~\ref{regimenes_massoft2}
and~\ref{regimenes_gsigma2} reflect the system dynamics in this
regime. Both mechanisms transfer energy from the expectation value
$ \eta $ to the quantum fluctuation $ \varphi_k $. Therefore, $
g\Sigma(\tau) $ increases while the amplitude of the oscillations
of $ \eta $ decreases.


\section{Intermediate time dynamics} \label{sec:inttime}

After the early period described in the previous section, the
system enters in a quasiperiodic regime. This behavior has been
already observed for the particular case $ T = 0 $ in Ref.
\cite{extfield}, while for different initial conditions the
oscillations damped much faster \cite{tsu, j0eta}

\begin{figure}
\begin{center}
\includegraphics [scale=0.7] {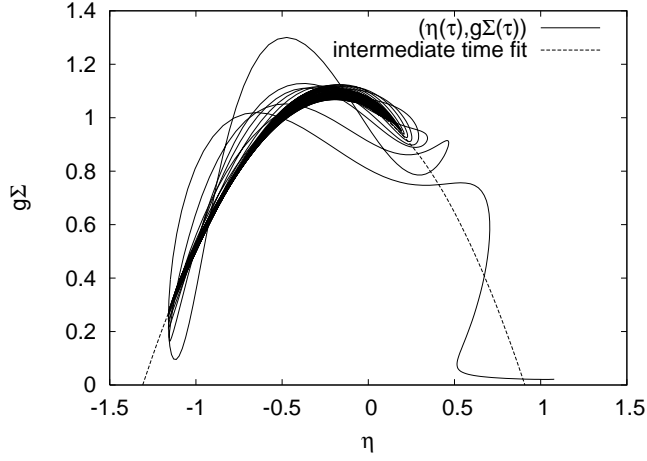}
\end{center}
\caption{Trajectory in the $ (\eta, g\Sigma) $ plane for
$\tau\leq1000$ and intermediate time ($ \tau \in [800,\,1000] $)
fit to Eq.~\eqref{Sigeta}, corresponding to the values
$g=10^{-2}$, $j=0.20$, $\beta^{-1}=1$.\label{gSeta}}
\end{figure}

This quasiperiodic behaviour found for $ \eta(\tau) $ and $
g\Sigma(\tau) $ suggests that these quantities are approximately
governed by an effective Hamiltonian with a few degrees of
freedom. Actually, we found (as in Ref. \cite{extfield}) that the
full numerical solution of Eqs.~\eqref{evolution1_adim}-\eqref{S}
gives that $ \eta(\tau) $ and $ g\Sigma(\tau) $ are approximately
related as

\be\label{Sigeta}
g\Sigma(\eta) = 1 + c_0 \, j - (1 - c_1 \, j)(\eta +j)^2
\ee
where $ c_0 $ and $ c_1 $ are positive numbers of the order $ j^0
$ and $ g^0 $ for small $ g $, that are obtained fitting the
numerical solution. See Fig.~\ref{gSeta}, and Tables~\ref{tc0} and
\ref{tc1}.

\bigskip

\begin{table}[h]
\begin{tabular}{ll|ccc}
\hline \hline
           & $ j  $ & $ 0.05 $ & $ 0.10 $ & $ 0.20 $ \\
$ \sqrt{g} \beta^{-1} $ &  &\multicolumn{2}{c}{ } \\
\hline
0   &  & $ 0.56 $ & $ 0.54 $ & $ 0.40 $  \\
0.1 &  & $ 0.48 $ & $ 0.48 $ & $ 0.37 $  \\
0.2 &  & $ 0.42 $ & $ 0.43 $ & $ 0.20 $  \\
0.3 &  & $ 0.40 $ & $ 0.40 $ & $ 0.11 $  \\
\hline \hline
\end{tabular}
\caption{ $ c_0 $ values obtained fitting $ g\Sigma(\eta) $ to
Eq.~\eqref{Sigeta} in the time interval $ \tau \in [800,\,1000] $
for $ g = 10^{-2} $ for various values of the external field $ j $
and the initial temperature. \label{tc0}}
\end{table}

\bigskip

\begin{table}[h]
\begin{tabular}{ll|ccc}
\hline \hline
           & $ j  $ & $ 0.05 $ & $ 0.10 $ &$ 0.20 $ \\
$ \sqrt{g} \beta^{-1} $ &  &\multicolumn{2}{c}{ } \\
\hline
0   &  & $ 1.49 $ & $ 1.02 $ & $ 0.55 $  \\
0.1 &  & $ 1.62 $ & $ 1.12 $ & $ 0.63 $  \\
0.2 &  & $ 2.03 $ & $ 1.42 $ & $ 0.98 $  \\
0.3 &  & $ 2.91 $ & $ 2.04 $ & $ 1.49 $  \\
\hline \hline
\end{tabular}
\caption{ $ c_1 $ values obtained fitting $ g\Sigma(\eta) $ to
Eq.~\eqref{Sigeta} in the time interval $ \tau \in [800,\,1000] $
for $ g = 10^{-2} $ for various values of the external field $ j $
and the initial temperature. \label{tc1}}
\end{table}

\bigskip

Using this approximate expression the effective squared mass can
be expressed as
\be  \label{M2etarel}
{\cal M}^2(\eta) = -1 + \eta^2 + g\Sigma(\eta) = j \left[ c_0 - j
+ c_1 j^2 -2( 1 - c_1 j )\, \eta + c_1 \, \eta^2 \right] \;.
\ee
The evolution equations (\ref{evolution1_adim})--(\ref{S}) in this
approximation take the form,
\be
\ddot \eta + {\cal M}^2(\eta)\, \eta = - j\;.
\ee
We find integrating on $ \eta $,
\be\label{efE}
\frac12 {\dot \eta}^2 + j \; V_{int}(\eta) = j\, E_{int}
\ee
where,
\bea
V_{int}(\eta) &=& \eta + \frac12 (c_0 - j + c_1 j^2)\eta^2
-\frac23( 1 - c_1 j )\, \eta^3 + {c_1 \over 4} \,\eta^4  \:, \\
E_{int} &=& V_{int}(\eta_1) \;.
\eea
Notice that  $E_{int}$ depends on the initial conditions, and that
$ \eta_1 $ is a turning point of the motion. Eq. (\ref{efE}) can
be integrated as follows,
\be\label{intel}
\sqrt{2 \, j}\,(\tau-\tau_1) = \int_{\eta_1}^\eta{
\frac{d\eta}{\sqrt{E_{int} -  V_{int}(\eta)}}}
\ee
with $ \eta_1 = \eta(\tau_1) $.

The fourth order polynomial $ E_{int} -  V_{int}(\eta) $ has
always two real roots $ \eta_1 < \eta_2 $ corresponding to the
turning points. Depending on the value of $ j $, the two other
roots can be:

\begin{itemize}
\item
i) a pair of complex conjugate roots $ \eta_R \pm i \eta_I $, for
which we define
\be
a \equiv \frac{\eta_R-\eta_1}{(\eta_R-\eta_1)^2+\eta_I^2} \; ,
\quad b \equiv \frac{\eta_I}{(\eta_R-\eta_1)^2+\eta_I^2} \; ,
\ee
\item
ii) or a pair of real roots $ \eta_1 < \eta_2 < \eta_3 < \eta_4 $,
we then define:
\be
a \equiv \frac{2}{\eta_3-\eta_1} - \frac{1}{\eta_4-\eta_1} \; ,
\quad b^2 \equiv {4 (\eta_3-\eta_2)(\eta_4-\eta_3) \over
(\eta_2-\eta_1)(\eta_4-\eta_1) (\eta_3-\eta_1)^2}\; .
\ee
\end{itemize}
We also introduce two  other quantities to simplify the formulas:
\be\label{cosas}
d \equiv \frac{1}{\eta_2-\eta_1} \; ; \quad X \equiv \left[
(a-d)^2 + b^2 \right]^{1/4} \; ; \quad C \equiv X\sqrt{2|{\cal
M}^2(\eta_1)\,\eta_1+j|}
\ee
It is convenient to use $u \equiv {\frac{1}{ \eta-\eta_1}}$ as
integration variable in Eq. (\ref{intel}). The solution of Eq.
(\ref{intel}) can be expressed after calculation as
\be  \label{etaanasol}
\eta(\tau) = \eta_1 +
\frac{(\eta_2-\eta_1)[1-\mbox{cn}(C(\tau-\tau_1),k)]}{1+(\eta_2-\eta_1)X^2
-[1-(\eta_2-\eta_1)X^2]\; \mbox{cn}(C(\tau-\tau_1),k)} \; ,
\ee
where $ \mbox{cn}(z,k) $ is the Jacobi cosine function, and
\be
k = \frac{1}{\sqrt{2}} \sqrt{1+\frac{a-d}{X^2}}
\ee
the elliptic modulus.

\begin{figure}
\begin{center}
\includegraphics [scale=0.7] {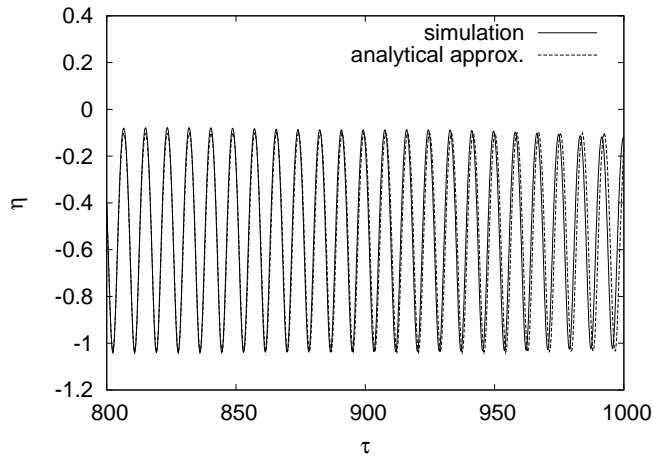}
\end{center}
\caption{Quantum evolution for the expectation value $ \eta(\tau) $ (full
  line) compared to the analytical intermediate time approximation
  of Eq.~\eqref{etaanasol} (dashed line). Corresponding to $ g=10^{-2} $,
  $ j=0.20 $ and $ \beta^{-1}=1 $ [for this case $j<j_c(\beta) $]. \label{jacobi}}
\end{figure}

The solution (\ref{etaanasol}) oscillates between $ \eta_1 $ and $
\eta_2 $
with period
\be
\mathcal{T} = \frac{4}{C}\, K(k)
\ee
where $ K(k) $ stands for the complete elliptic integral of the
first kind and $ C $ is given by Eq. (\ref{cosas}). The analytical and the
numerical solutions are compared in Fig. \ref{jacobi}.

As in the zero temperature case (Ref.~\cite{extfield}), the
numerical solution of the large $ N $ limit evolution equations
\eqref{evolution1_adim}-\eqref{S} would be periodic if $ E_{int} $ were exactly
conserved; however it is slowly decreasing implying a slow damping
of the oscillations. This damping is smaller the smaller is $j$.
See Table~\ref{Eint}.

\begin{table}[h]
\begin{tabular}{c|rrr}
\hline \hline
$ \tau $ & $ 400 $ & $ 700 $ & $ 1000 $ \\
\hline
 $ jE_{int}(j=0.05,\,\sqrt{g}\beta^{-1}=0.0 ) $
   & $ 0.022 $ & $ 0.016 $ & $ 0.015 $  \\
 $ jE_{int}(j=0.05,\,\sqrt{g}\beta^{-1}=0.1 ) $
   & $ 0.019 $ & $ 0.013 $ & $ 0.009 $  \\
 $ jE_{int}(j=0.05,\,\sqrt{g}\beta^{-1}=0.2 ) $
   & $ 0.015 $ & $ 0.011 $ & $ 0.006 $  \\
 $ jE_{int}(j=0.05,\,\sqrt{g}\beta^{-1}=0.3 ) $
   & $ 0.012 $ & $ 0.007 $ & $ 0.006 $  \\
\hline
 $ jE_{int}(j=0.20,\,\sqrt{g}\beta^{-1}=0.0 ) $
   & $ 0.011 $ & $ 0.004 $ & $ -0.011 $  \\
 $ jE_{int}(j=0.20,\,\sqrt{g}\beta^{-1}=0.1 ) $
   & $ -0.004 $ & $ -0.014 $ & $ -0.024 $  \\
 $ jE_{int}(j=0.20,\,\sqrt{g}\beta^{-1}=0.2 ) $
   & $ -0.027 $ & $ -0.035 $ & $ -0.046 $  \\
 $ jE_{int}(j=0.20,\,\sqrt{g}\beta^{-1}=0.3 ) $
   & $ -0.050 $ & $ -0.058 $ & $ -0.064 $  \\
\hline \hline
\end{tabular}
\caption{ $ E_{int} $ [Eq.~\eqref{efE}] for $ g = 10^{-2} $ and
various values of the external field $ j $ and the initial
temperature. \label{Eint}}
\end{table}

Therefore, we have seen that the system at intermediate times
presents a clear separation between fast variables and slow
variables. $
\eta(\tau) $, $ g\Sigma(\tau) $, and $ {\cal M}^2(\tau) $
oscillate fast, while $ E_{int} $ slowly decreases. We have found
the fast time dependence of the order parameter and the quantum
fluctuations in terms of rational functions of Jacobi cosines.

The analytic expression for $ g\Sigma(\tau) $, and $ {\cal
M}^2(\tau) $ are obtained from the relations $ g\Sigma(\eta) $
[Eq. (\ref{Sigeta})] and $ {\cal M}^2(\eta) $  [Eq.
(\ref{M2etarel})], just using the analytical solution for $
\eta(\tau) $.
The relation $ g\Sigma(\eta) $ indicates that $ g\Sigma $ and $
\eta $ oscillate with opposite phase, up to order $ j $ terms; $
g\Sigma(\eta) = 1 - \eta^2 + {\cal O}(j) $. This $ {\cal O}(j) $
terms are very important because they imply $ {\cal M}^2(\tau) =
-1 + \eta^2 + g\Sigma = {\cal O}(j) $. Therefore the time averaged
squared mass is positive, and tends asymptotically to a positive
value. This can be seen also from Eq.~(\ref{M2etarel}). Thus, in
the limit of zero external field we recover a zero effective mass
squared consistently with the presence of Goldstone bosons for
zero external field. [Recall that for $ j=0 $ in the broken
symmetry case the squared mass goes to zero for initial energies
below the potential at the local maximum ($\eta = 0 $)
\cite{j0eta,tsu}.]

\section{Conclusions}

We have studied here the effects of the initial temperature in the
out of equilibrium dynamics induced by a flip in the sign of the
external field. The study has been done in the leading order of
large $ N $ for the $ \lambda(\vec\Phi^2)^2 $ theory.

Before the flip of the external field, the system is in a thermal
state, with the expectation value of the field in the minimum of
its effective potential. After the flip the system evolves in
either of two dynamical regimes. The difference between the two
regimes is the presence or absence of a temporal trapping of the
system in the neighborhood of a metastable state. The presence of
this trapping close to a metastable state is reflected in the fact
that the expectation value oscillates without changing sign
(despite of the sign change of the external field).

From the evolution equation in the large $ N $ limit, we have
found a dynamical effective potential for the expectation value in
the direction of the external field that helps to understand the
dynamics. The trapped regime corresponds to the case where the
system cannot overcome the barrier appearing in the dynamical
effective potential. We have shown that the dynamical regime is
determined by the initial temperature and the external field.
There is a temperature dependent critical value of the external
field that separates both regimes, see Eq.~\eqref{jcbeta}. For
external fields smaller than the critical one, the system is
temporally trapped close to the metastable equilibrium position of
the effective potential.
For external fields greater than the critical one, the neighborhood of
the stable equilibrium is rapidly reached and no trapping is
observed.

Spinodal growth of the quantum fluctuations is the main mechanism
that makes the trapping be temporal. The increase of the quantum
fluctuations diminishes the potential barrier between the two
minima of the dynamical effective potential for the expectation
value. We have found an approximate analytic expressions for the
trapping time (spinodal time), see Eq.~\eqref{tsfinal}.
It shows that the trapping time becomes shorter for
larger initial temperatures or larger external fields.

After the trapping ends, the transfer of energy from the
expectation value to the quantum fluctuations takes place through
two mechanism: spinodal instabilities and parametric resonances.
The spinodal instabilities appear during the time intervals when
the effective mass is negative, while the parametric resonances
are due to the oscillations of the expectation value.

We have found that these processes lead for intermediate times to
a quasiperiodic regime with a clear separation of slow and fast
variables. We have explicitly solved the time evolution for the
fast variables that oscillate periodically, while the other
variables change slowly due to a tiny damping of the oscillations.
The effective squared mass oscillates around a positive value of
the order of magnitude of the external field. Thus, in the limit
of zero external field we recover a zero effective mass squared
consistently with the presence of Goldstone bosons for zero
external field \cite{tsu}.

An interesting question that remains still open is which of these
results would be modified by the next to leading order terms of
the large $ N $ approximation \cite{berges} and how much. The
inclusion of next to leading order terms allows to include the
effects of longitudinal fluctuations. (The longitudinal
fluctuations have quantum and thermal origin, and later they
reflect the effects of thermal jumping over the barrier and of
quantum tunnelling through the potential barrier.) This inclusion
will allow to compute the probability of finding the system close
to the stable minimum before the spinodal time, and it will also
give corrections to the spinodal time; these corrections are
expected to be small, as the spinodal time will be mainly
determined by the quasiexponential growth of the transversal
fluctuations. Also the inclusion of next to leading order terms
have been shown to damp faster the oscillations on the magnitudes
during evolution \cite{berges}. Therefore, how much the next to
leading order terms will increase the damping rate of the
intermediate time quasiperiodic behavior is still an open
question.

\acknowledgments

We thank Hector J. de Vega for useful comments.
We acknowledge financial support from the Ministerio de Ciencia y
Tecnolog\'{\i}a of Spain through the Research Project
BFM2003-02547/FISI.

\appendix
\section{Spinodal time}\label{tiempoespinodal}
We present here the early time solution for the modes in the
spinodally resonant band, and an estimate for the spinodal time, $
\tau_s $, in the case where the external field is smaller than the
critical value, $j<j_c(\beta)$. In this case the spinodal time
corresponds to the time the system is trapped close to the
metastable state.

At early times (before the spinodal time, $\tau_s$), the effective
squared mass, ${\cal M}^{2}(\tau)$, oscillates with a negative
average. Thus, we can define an effective average squared mass
$-\mu^2$ and replace ${\cal M}^{2}(\tau)$ by
$-\mu^2$ in the modes evolution
equation~\eqref{evolution1_adim} to obtain an estimate for $|\varphi_q|$. This
is a good
approximation when the time dependence of ${\cal M}^{2}(\tau)$ is
slow (adiabatic approximation), and after several complete
oscillations of ${\cal M}^{2}(\tau)$. Assuming $ g\Sigma(\tau) \ll
1 $ (otherwise $\tau_s$ is very short), we have ${\cal
M}^2(\tau)\approx -1+\eta^2(\tau)$. As $\eta$ evolves according to
Eq.~\eqref{evolution1_adim} with a frequency ${\cal M}(\tau)$,
both $\eta$ and ${\cal M}^2(\tau)$ oscillate with a period of the
same order $\frac{1}{\mu}$. Therefore, as the effective average squared
mass approximation is valid when we average over several periods,
this requires
\begin{equation}\label{periodo}
\tau\gtrsim\frac{1}{\mu}\; .
\end{equation}
(We will show at the end of this appendix that requiring $ \tau_s
> 1/\mu $ implies $ g\Sigma(\tau) \ll 1 $ during most of the time
interval $ \tau \in [0,\, \tau_s] $.)

The average squared mass, $ -\mu^2 $, is estimated as the average
value of ${\cal M}^2$ between its maximum ${\cal M}_{max}^2$, and
minimum ${\cal M}_{min}^2$ value. These values correspond to the
$\eta$ field oscillations between $\eta_0$ [Eq.~\eqref{eta0}], and
the turning point value $\eta_r$ [Eq.~\eqref{etar}]. Lets call
$\tau_P$ the time over which the average is done, it results
\begin{equation}\label{-mu2}
-\mu^2=\frac{1}{\tau_P}\int_0^{\tau_P} {\cal M}^2(\tau)\;d\tau\simeq
  \frac{1}{2} \left({\cal
  M}_{min}^2 + {\cal M}_{max}^2\right);
\end{equation}
Neglecting $g\Sigma$ for these times $\tau<\tau_P$,
\begin{equation}\begin{split}
{\cal M}_{max}^2 &\approx -1+\eta_0^2\approx j-\frac{1}{2}j^2
+\frac{5}{8}j^3-j^4+{\cal O}(j^5),\\
{\cal M}_{min}^2 &\approx -1+\eta_r^2\approx -3j-\frac{1}{2}j^2
-\frac{23}{8}j^3-4j^4+{\cal O}(j^5).
\end{split}\end{equation}
Using these expansions on $j$ we get from the Eq.~\eqref{-mu2}
\begin{equation}\label{-mu2enj}
-\mu^2=-j-\frac{1}{2}j^2-\frac{9}{8}j^3-\frac{5}{2}j^4+{\cal O}(j^5).
\end{equation}
\par
The evolution equations for the modes are then
\begin{equation}
\left( \frac{d^2}{d\tau^2}+q^2-\mu^2\right)\varphi_q(\tau)=0
\end{equation}
with the initial conditions
\begin{equation}
\varphi_q(0)=(q^2+{\cal M}^2(0))^{-1/4}\;,\quad
\dot\varphi_q(0)=-i(q^2+{\cal M}^2(0))^{1/4}\;.
\end{equation}
For low momenta (those in the band $0\leq q\leq \mu$) the effective
squared frequency $q^2-\mu^2$ is negative (spinodal instability)
and the $\varphi_q(\tau)$ have an exponential behavior.
\par
Solving this second order ordinary differential equation,
\begin{equation}\begin{split}\label{modos}
\varphi_q(\tau)&=\frac{1}{2\sqrt{\mu^2-q^2}(q^2+{\cal M}^2(0))^{1/4}}\\
&\quad\times\bigg[\left( \sqrt{\mu^2-q^2}-i\sqrt{q^2+{\cal
M}^2(0)}\right) e^{\tau\sqrt{\mu^2-q^2}}\\
&\quad+\left( \sqrt{\mu^2-q^2}+i\sqrt{q^2+{\cal M}^2(0)}\right)
e^{-\tau\sqrt{\mu^2-q^2}}
\bigg]\;.
\end{split}\end{equation}
The contribution of the spinodal band to $g\Sigma$ is
\begin{equation}\label{gsigma}
g\Sigma_s(\tau)=g\int_{0}^{\mu}{dq\,q^2\vert\varphi_q(\tau)\vert ^2
\coth{\left( \frac{\beta\omega_q(0)}{2}\right)}}\;.
\end{equation}
This spinodal band gives the dominant contribution to
$g\Sigma_s(\tau)$. The exponential growth of these low momenta modes
makes that after a short time they give the main contribution.
\par
Inserting Eq.~\eqref{modos} in Eq.~\eqref{gsigma} we have
\begin{equation}
g\Sigma_s(\tau)=g\int_{0}^{\mu}{dq\, \biggl\{ q^2\frac{\mu^2+{\cal
M}(0)^2}{4\mu^2\left( 1-\frac{q^2}{\mu^2}\right)\sqrt{q^2+{\cal
M}(0)^2}}} \; e^{2\tau\mu\sqrt{1-\frac{q^2}{\mu^2}}} \coth{ \left[
\frac{\beta\sqrt{q^2+{\cal M}(0)^2}}{2} \right] }
\biggr\}\;,
\end{equation}
where we have neglected the exponentially decreasing term of
$\varphi_q$. This is justified by the condition~\eqref{periodo}.

It is convenient to separate three dynamical regimes:
\bi
\item High temperatures, $ \beta^{-1} \gg \mu/g $, this
implies $ g\Sigma(0) \gtrsim 1 $, i.e., $ j \gtrsim j_c(\beta) $, and
there is no trapping.
\item Medium temperatures, $ \mu/g \gg \beta^{-1} \gg \mu $, this
implies $ g\Sigma_s(\tau) \approx \frac{2}{\beta\mu}
g\Sigma_s^{T=0}(\tau) $ as we show later.
\item Low temperatures, $ \beta^{-1} \ll \mu $, this
implies $ g\Sigma_s(\tau) \approx g\Sigma_s^{T=0}(\tau) $ as we
show later.
\ei
Therefore, here we are only interested in the medium and low
temperature cases where the system can be trapped close to the
metastable state giving a nonzero spinodal time. We devote the
following subsections to these cases.

\subsection{Medium temperatures ($ \mu/g \gg \beta^{-1} \gg \mu $)}
We call medium temperatures those that verify
$\beta^{-1}\gg\omega_q(0)$ for all the modes in the spinodal band,
i.e.,
\begin{equation}
\beta^{-1}\gg\mu\;;
\end{equation}
and also give $ g\Sigma(0) \ll 1 $, that implies
\be
\beta^{-1}\ll\mu/g\;.
\ee
This allows the following simplifications
\begin{equation}
{\cal M}^2(0)=-1+\eta^2(0)+g\Sigma(0)\approx -1+\eta^2(0)\;,
\end{equation}
and also in this regime ($\beta^{-1}\gg\mu$)
\begin{equation}\label{cothapprox}
\coth{\left( \frac{\beta\omega_q(0)}{2}\right)}= \coth{\left(
\frac{\beta\sqrt{q^2+ {\cal
M}^2(0)}}{2}\right)}\approx\frac{2}{\beta\sqrt{q^2+{\cal
M}^2(0)}}\;.
\end{equation}
Therefore,
\begin{equation}
g\Sigma_s(\tau) = g\int_{0}^{\mu}{dq\, \biggl\{
q^2\frac{\mu^2+{\cal M}(0)^2}{4\mu^2\left(
1-\frac{q^2}{\mu^2}\right)\sqrt{q^2+{\cal M}(0)^2}}} \;
e^{2\tau\mu\sqrt{1-\frac{q^2}{\mu^2}}}
\frac{2}{\beta\sqrt{k²+{\cal M}^2(0)}} \biggr\} \;.
\end{equation}
In addition, ${\cal M}^2(0)$ is of the same order of magnitude as
$\mu^2$, and we make the approximation ${\cal M}^2(0)\approx
\mu^2$. We get
\begin{equation}
g\Sigma_s(\tau)=\frac{g}{\beta} \int_{0}^{\mu}{dq\,
q^2\frac{1}{\mu^2\left( 1-\frac{q^2}{\mu^2}\right)\left(
1+\frac{q^2}{\mu^2}\right)}} \;
e^{2\tau\mu\sqrt{1-\frac{q^2}{\mu^2}}} \;.
\end{equation}
It can be shown that the greatest contributions to the integral
come from the integrand values that have $q={\cal O}(0.1\mu)$ (the
exponential enhance low momenta while other factors suppress very
low momenta). This justifies making a Taylor expansion to second
order in $\frac{q}{\mu}$, both in the exponential and in the
previous factor. The result obtained is
\begin{equation}
g\Sigma_s(\tau)=\frac{g}{\beta} \int_{0}^{\mu}{dq\,
\frac{q^2}{\mu^2} \;
e^{2\tau\mu\left(1-\frac{q^2}{2\mu^2}\right)}} \;,
\end{equation}
and making the change of variable $\xi=q\sqrt{\frac{\tau}{\mu}}$
the previous expression becomes
\begin{equation}
g\Sigma_s(\tau)=\frac{g}{\beta}
\frac{e^{2\tau\mu}}{\sqrt{\mu}\tau^{3/2}}
\int_{0}^{\sqrt{\mu \tau}}{d\xi\, \xi^2 e^{-\xi^2}}
\;.
\end{equation}
For $\mu\tau\to \infty$ the integral gives $\sqrt{\pi}/4$, that is
a good approximation for $\mu \tau\gg 1$
[condition~\eqref{periodo}]. Therefore, a good approximate
expression for the growth of the quantum and thermal contributions
in the spinodal band is
\begin{equation}\label{gsigmaT}
 g\Sigma(\tau)\approx
\frac{g}{4\beta}\sqrt{\frac{\pi}{\mu}}
\frac{e^{2\tau\mu}}{\tau^{3/2}}
\;,
\end{equation}
that is valid for times $\frac{1}{\mu}\lesssim \tau\lesssim
\tau_s$ and temperatures $\mu\ll \beta^{-1}\ll\frac{\mu}{g}$.
\par
In the case of zero temperature we have
$g\Sigma_s^{T=0}(\tau)\approx\frac{g\sqrt{\pi\mu}e^{2\tau\mu}}{8\tau^{3/2}}$
(see Ref.~\cite{extfield}) and we can express the result obtained
as
\begin{equation}
g\Sigma_s(\tau)\approx\frac{2}{\beta\mu}g\Sigma_s^{T=0}(\tau)\; .
\end{equation}
After a certain time, the spinodal time $\tau_s$, the quantum and
thermal effects start to be important in the dynamics,
$g\Sigma_s(\tau_s)$ compensates $-\mu^2$ and the exponential
growth of the mode functions stops, and they start to all have an
oscillatory behavior. Thus, the spinodal time $\tau_s$ is defined
as
\begin{equation}\label{ts}
g\Sigma_s(\tau_s)=\mu^2\;.
\end{equation}
Using Eqs.~\eqref{gsigmaT} and~\eqref{ts} we have
\begin{equation}
\frac{g}{4\beta}\sqrt{\frac{\pi}{\mu}}
\frac{e^{2\tau_s\mu}}{\tau_s^{3/2}}=\mu^2
\;,
\end{equation}
and taking logarithms we obtain
\begin{equation}
\tau_s=\frac{1}{2\mu}
\left[
\log{\left( \frac{4\beta\mu}{g\sqrt{\pi}} \right)}+
\frac{3}{2}\log{\left(\mu \tau_s\right)}
\right]
\;,
\end{equation}
implying
\begin{equation}\begin{split}\label{tausT}
\tau_s &=
\frac{1}{2\mu}
\log{\left( \frac{4\beta\mu}{g\sqrt{\pi}} \right)}+
\frac{3}{4\mu}\log{\left[\frac{1}{2}
\log{\left( \frac{4\beta\mu}{g\sqrt{\pi}} \right)}
\right]}
\\
&\quad +{\cal O}\left(
\frac{\log\log\frac{1}{g}}{\log\frac{1}{g}}
\right)
\;;
\end{split}\end{equation}
this expression is related with the spinodal time for zero
temperature, $\tau_s^{T=0}$, (see Ref.~\cite{extfield}) through
the equation
\begin{equation}\label{tsT}
\tau_s=\tau_s^{T=0}-\log{\left( \frac {2}{\beta\mu} \right)}+{\cal O}\left(
\log\log\frac{1}{g}
\right)
\; .
\end{equation}
One direct conclusion from the expression~\eqref{tsT} is that an
increase of the initial temperature diminishes the spinodal time.

\subsection{Low temperatures ($ \beta^{-1}\ll\mu $)}
For temperatures that satisfy
\begin{equation}
\beta^{-1}\ll\mu
\end{equation}
we can make the following development for the hyperbolic
cotangent,
\begin{equation}
\coth{\left( \frac{\beta\omega_q(0)}{2}\right)}=
\coth{\left( \frac{\beta\sqrt{q²+{\cal M}^2(0)}}{2}\right)}=
1+{\cal O}\left(e^{-\beta\mu} \right)\;.
\end{equation}
Thus, for these low temperatures we recover as a good
approximation the zero temperature expressions, and we obtain for
the spinodal time a value very close to $\tau_s^{T=0}$.
\par
\bigskip
In summary, a good approximate expression for the spinodal time to
order ${\cal O}\left( \frac{\log\log\frac{1}{g}}{\log\frac{1}{g}}
\right)$ is
\begin{equation}
\tau_s=
\begin{cases}
\frac{1}{2\mu}
\log{\left( \frac{4\beta\mu}{g\sqrt{\pi}} \right)}+
\frac{3}{4\mu}\log{\left[\frac{1}{2}
\log{\left( \frac{4\beta\mu}{g\sqrt{\pi}} \right)}
\right]}&\text{for $\beta^{-1}\gg\mu$} \;,\vspace{1mm}\\
\frac{1}{2\mu} \log{\left( \frac{8}{g\sqrt{\pi}} \right)}+
\frac{3}{4\mu}\log{\left[\frac{1}{2} \log{\left(
\frac{8}{g\sqrt{\pi}} \right)} \right]}& \text{for
$\beta^{-1}\ll\mu$} \;.
\end{cases}
\end{equation}
This analytic expression is in good agreement with the numerical results,
as can be seen in Table~\ref{tabla}.

\end{document}